\documentclass[12pt,preprint]{aastex}
\begin{document}

\title{A Survey for Young Spectroscopic Binary K7$-$M4 Stars in Ophiuchus}

\author{L. PRATO\altaffilmark{1} }

\altaffiltext{1}{Lowell Observatory, 1400 West Mars Hill
  Rd. Flagstaff, AZ 86001; lprato@lowell.edu}

\begin{abstract}

This paper describes a high-resolution, infrared spectroscopic
survey of young, low-mass stars designed
to identify and characterize pre$-$main-sequence
spectroscopic binaries. This is the first large
infrared radial velocity survey of very young stars to
date. The frequency and mass ratio distribution
of the closest, low-mass binaries bear directly on models of stellar,
brown dwarf, and planetary mass companion formation. Furthermore,
spectroscopic binaries can provide mass ratios and ultimately
masses, independent of assumptions,
needed to calibrate models of young star evolution.
I present the initial results from observations of a uniform sample
of 33 T Tauri M stars in the Ophiuchus molecular cloud.
The average mass of this sample is less than that of other young
star radial velocity surveys of similar scope by a factor of $\sim$2.
Almost every star was observed at 3$-$4 epochs over 3 years with the
10 meter Keck II telescope and the facility infrared spectrometer
NIRSPEC.  An internal precision of 0.43~km~s$^{-1}$ was obtained with
standard cross-correlation calibration techniques.
Four of the targets are newly discovered
spectroscopic binaries, one of which is located in a sub-arcsecond,
hierarchical quadruple system.  Three other sub-arcsecond visual
binaries were also serendipitously identified during target
acquisition.  The spectroscopic
multiplicity of the sample is  comparable to that of earlier type,
pre$-$main-sequence objects.  Therefore, there is no dearth of young,
low-mass spectroscopic binary stars, at least in the Ophiuchus
region.

\end{abstract}


\section{Introduction}

The ratio of binary to single stars on main-sequence is known to vary as
a function of spectral type (Abt 1983).  In some star forming
regions (SFRs), this ratio is known to be in excess
of the field star binary frequency (Duquennoy \& Mayor 1991), however, in
other SFRs, it is the same as that of the field stars.
Siegler et al. (2003, 2005) and Burgasser et al. (2003)
find evidence for a decreasing wide binary frequency associated
with very low-mass stars and brown dwarfs in the field.
Binary frequency appears to also be a
function of the mass ratio, as illustrated, for example, by
the differences in the numbers of stellar versus substellar
companions to G stars (e.g., Duquennoy \& Mayor 1991; Mazeh
et al. 1992; Marcy \& Butler 2000).
As searches for binary systems have become more complete,
the observational findings create ever more stringent
conditions that must be accounted for in any uniform model of star,
brown dwarf, and planet formation.  However, because the extent of
dynamical evolution is an open question,
the most useful observations for constraining binary
formation models are those based on the youngest observable stars.

Multiplicity surveys of the last decade and a half have shown
a surprising diversity of visual binary frequency in nearby star forming
regions (Ghez et al. 1993;
Leinert et al. 1993; Simon et al. 1995; Patience et al. 2002).
Taurus is particularly rich in multiples, however,
other regions appear to have solar neighborhood stellar companion
frequency (Petr et al. 1998; Beck et al. 2003; Ratzka et al. 2005).
The differences appear to result from the influence of initial
conditions in these regions on the frequency of binary formation
(Mathieu et al. 2000; Sterzik et al. 2003; K\"ohler et al. 2006).
The frequency of spectroscopic binaries in the same regions which
have been so well-studied for visual binaries is not as well-known.
The early, 1 to a few~km~s$^{-1}$ precision
radial velocity measurements of Herbig (1977),
Hartmann et al. (1986), Walter et al. (1988; 1994),
Mathieu et al. (1989), and Mathieu (1994) identified the largest samples of 
spectroscopic binaries in the Taurus and Ophiuchus star forming
regions published to date.  Recent radial velocity surveys of young
stars (e.g., Guenther et al. 2001; Melo 2003; Huerta et al. 2005;
Covey et al. 2006; Joergens 2006; Kurosawa et al. 2006) for stellar and
substellar companions are just
beginning to produce results.  Some of these next generation surveys
are larger in scope and some are higher in precision than previous
work.  Large, complete surveys drawn from 
consistent samples are not yet available but will provide extremely
useful input for theorists studying binary and planet formation
(e.g., Delgado-Donate et al. 2004; Boss 2006).

Besides extending our knowledge of young star multiplicity to very
low masses, spectroscopic binaries are important in their own right.
Such systems allow precise determination of dynamical
properties such as the mass function (for single-lined systems, SB1s),
the mass ratio (for double-lined systems, SB2s), and ultimately
component masses (for SB2s which are eclipsing or have angularly
resolved orbits, yielding orbital inclination).
Mass is the single most important stellar property from which,
given the metallicity, all other properties of an object may be
derived.  Accurate masses are necessary to calibrate the theoretical
pre$-$main-sequence (PMS) evolutionary tracks (Simon et al. 2000;
Hillenbrand \& White 2004), particularly at low masses where there
is significant scatter in models and very few objects with known
masses (White et al. 1999; Prato et al. 2002a; Torres \& Ribas 2002;
Stassun et al. 2006).  Furthermore,
mass ratio distributions and secondary mass distributions provide
additional input to theories of binary formation.  Thus, 
the identification of young spectroscopic binaries is extremely useful.
Young M stars are of particular interest in that they are
the most common stellar spectral type in most star
forming regions, including Ophiuchus, regardless of where exactly
the mass function turns over near the substellar boundary
(e.g., Luhman et al. 2000).

There are very few young, low-mass SB2s known.  Only 12.9\% of the
31 SB2s listed in Prato et al. (2002a) have primary masses less
than 1 solar mass.  Although additional PMS SB2s have been
discovered since 2002, the small percentage of sub-solar mass
systems prevails.
The reason for this bias is natural: surveying low-mass stars
in SFRs with minimum distances of $\sim$140~pc,
and therefore relatively faint magnitudes,
is impractical at 1$-$3~m class telescopes where most of the radial
velocity studies of young stars have been conducted with
visible light spectrometers.  Furthermore,
radial velocity surveys for long-period systems, or for systems with
very low-mass companions, require long time series observations.
The 8$-$10~m class telescopes have only been fully equiped and
commissioned in the last 5$-$10 years, therefore work of this nature has
only recently become possible.

My colleagues and I have developed a powerful approach to the
study of SBs using high-resolution infrared spectroscopy (Prato
1998; Steffen et al. 2001; Prato et al. 2002a, b; Mazeh et al.
2002; 2003; Simon \& Prato 2004).  Low-mass primary stars
emit most of their radiant energy at wavelengths $>$1~$\mu$m.
Also, because luminosity scales as a steeper function of mass
in the $V$-band than, for example, in the $H$-band,
it is easier to detect faint, low-mass secondary stars in
the blended spectra of SBs at longer wavelengths.  Finally,
the lower amplitude of star spot
modulated fluctuations in the infrared compared to visible light
also favors longer wavelength observations (Carpenter et al. 2001).
It is simply most efficient to search
for young M star spectroscopic binaries with a large aperture telescope
in the infrared.

This paper describes the initial results from a radial velocity
survey of a homogeneous sample of M stars in the Ophiuchus
SFR with infrared observations at
the Keck II 10-m telescope.  The goal was to find young
M star spectroscopic binaries in Ophiuchus by searching for
radial velocity variability and to measure the mass ratios
of any new systems discovered.  Two SB2s, two SB1s, and
one radial velocity variable were found.  The results presented are
preliminary in this first report; because the phase coverage
is sparse, it is only possible at this point to measure mass ratios.
Determination of the periods and other orbital elements will require
intensive followup.  A natural outcome of observing with a large
aperture telescope located at a site with good seeing was the
additional, serendipitous discovery of a number of close, visual
binaries.  In \S2 I describe the sample selection and the template
star library used for radial velocity measurements.  The observations
and data reduction are detailed in \S3.  The
results are presented in \S4.  \S5 provides a discussion
and a summary appears in \S6.

\section{Sample Selection and Spectral Type Templates}

All the targets were culled from the spectroscopic classification
paper of Mart\'{i}n et al. (1998).  The objects in that paper were
selected for association with $ROSAT$ X-ray sources, i.e. location
within 30$''$, and optical brightness, i.e. V$\le$15~mag.
The {\it 2MASS} $H$-band
magnitudes of my sample range from 8.6 to 11.1 and the spectral
types, as presented in Mart\'{i}n et al., from K6 to M4.5 (but see \S4.1).
This is not a complete sample; seven K6 and K7 stars and five
M stars from Mart\'{i}n et al. were not observed.  However, it is
a homogeneous sample in that all of the targets are X-ray sources,
are located in Ophiuchus, and are young.
Mart\'{i}n et al. determined youth and membership with
Li~6708\AA~measurements.  

I originally chose 31 M star targets in 28 fields; several fields
around the coordinates given in Mart\'{i}n et al. (1998) contained
multiple candidates.  In one such field (RX~J1625.2$-$2455),
two spectra taken of the fainter object, ``b'' in Table 1 of
Mart\'{i}n et al., revealed a large radial
velocity ($\sim$130~km~s$^{-1}$).  This object was subsequently excluded
from the sample as a probable halo star.  It is likely the brighter
``a'' component that is associated with the large Li equivalent
width in Table 2 of Mart\'{i}n et al..  Both components in three of the
subarcsecond, visual binaries, discovered in this study, were observed
individually.  Thus the final number of unresolved objects surveyed in this
paper is 33, located in 30 distinct systems.  Four of these
are classical T~Tauri stars (CTTS), 24
are weak-lined T~Tauri stars (WTTS), and 5 are post-T~Tauri stars (PTTS),
according to the classification of Mart\'{i}n et al.  Among the
distinct regions within Ophiuchus described in Mart\'{i}n et al.,
I selected 2 targets from the ``streamers'', 12 from the
``$\rho$ Oph core'', 7 from ``R7'', and 12 from the ``smoke
rings'' (see Mart\'{i}n et al. Figure 1).  The target names
appear in Table 1, column (1).  Accurate coordinates for the objects
were determined by iterating and refining the initial coordinates
using the $2MASS$ interactive image service; these appear in columns
(2) and (3) of Table 1.  If two relatively bright stars appeared
in the same field near the coordinates given by Mart\'{i}n et al.
and were both observed, they are designated with the same name
followed by a letter indicating cardinal orientation.  The PMS
evolutionary class, CTTS (C), WTTS (W), or PTTS(P), from
Mart\'{i}n et al. is given in column (4) of Table 1.

A subset of the spectral type templates from Prato et al. (2002a) and
Bender et al. (2005) were used in this current study to measure
radial velocities.  Figure 1 of Prato et al. shows the spectral
types of these templates as given in SIMBAD.  The FeI and OH
features at $\sim$1.563$\mu$m are especially useful in the
determination of spectral types.  In order to assign spectral types
to the targets and to thereby determine which templates to use for
radial velocity measurements, I reordered the sequence of 
M0 to M4 templates to show a logical progression of FeI and OH
features.  T. Barman kindly produced a synthetic sequence
of Phoenix model spectra from the NexGen grid (Hauschildt et al. 1999)
over a similar temperature range for comparison.
I assigned spectral types to the M type template stars on the basis
of their atomic and molecular features and used the synthetic
sequence as a guide (Figure 1 and Table 2).  Interestingly,
with the exception of the first (M0) and last (M4) objects in this series,
the best progression that I identified differs from the
spectral types for these standards adopted in the literature, based
primarily on low- to medium-resolution visible light spectroscopy.
Prato et al. (2002b) also note this mismatch.
This discrepancy is likely the result of a number of factors.
The visible and infrared light spectra do not necessarily sample
the same regions in the M star atmospheres.  Furthermore, at the
high-resolution ($\sim$30,000) of the infrared observations,
the spectra are likely affected by subtle differences in atmospheric
mixing, metallicity, surface gravity, etc.
(J. D. Kirkpatrick, priv. comm.).

\section{Observations and Data Reduction}

All data were obtained at the Keck~II 10-m telescope on
Mauna Kea using NIRSPEC, the facility near-infrared,
cross-dispersed, cryogenic spectrometer (McLean et al. 1998; 2000).
NIRSPEC employs a 1024 $\times$ 1024 ALADDIN InSb array detector.
I observed in the $H$-band, at a central wavelength of $\sim$1.555$\mu$m,
with an 0.288$''$ (2 pixel) $\times$ 24$''$ slit, yielding R$=$30,000.
Dispersion solutions were derived from OH night sky emission lines
for all spectra; these are very well distributed across order 49,
the central order at this setting.  Another advantage of order 49 is
the relative lack of telluric absorption lines in the region.
Consequently, I did not divide out telluric standard spectra and
used only order 49 in the subsequent analysis.
Source aquisition was accomplished with the slit viewing camera,
SCAM, which utilizes a 256 $\times$ 256 HgCdTe detector with
0.18$''$ pixels.

The observations were made between March, 2002, and July, 2004.
$H$-band magnitudes for the sample range from 8.6 to 11.1; total
integration times depended on source brightness and seeing and
typically varied from 10$-$40 minutes, in individual frames of
100$-$300~s.  Three to four spectra were taken for almost all of the
targets over the 3 years of observations.  For one star, only
one spectrum was obtained; for another,
only two spectra were obtained.  The $2MASS$ $H$-band magnitudes for the
targets and the number of NIRSPEC observations are listed in columns (5)
and (6) of Table 1.

The REDSPEC package, software designed at UCLA by S. Kim, L. Prato,
and I. McLean specifically
for the analysis of NIRSPEC data\footnote{See: 
http://www2.keck.hawaii.edu/inst/nirspec/redspec/index.html},
was used for all reductions.  The observational set up and data
reduction procedures followed were identical to those described
in Prato et al. (2002a).  This is true for both target stars as
well as the spectral type template stars used in the analysis here.
Although most of the template stars were observed on runs
in 2000$-$2001 during which NIRSPEC was used in conjunction
with the Keck II adaptive optics (AO) system (Prato et al.;
Mazeh et al. 2003), this does not introduce any
significant differences in the resultant spectra.  The same is true
for the one target spectrum taken with NIRSPEC behind the AO system.

For the three visual binaries (\S 4.2)
discovered during the course of these
observations which had the angularly resolvable separations of 0.7$''$ to
1.0$''$ in conditions of good seeing (0.4$-$0.8$''$), both objects
were usually placed on the appropriately rotated slit and observed
simultaneously.  For RX J1612.6$-$1924E, only
once were both components observed in this way.
The traces of the wider pairs were easily
separable.  For the closest visual pair, narrow apertures were used
to sum the flux from each component, on the outside extremes of the
spectral traces in order to avoid contamination of each spectrum
by the other.  A cross-cut of the trace of an apparently single
target was used to judge the extent of the overlapping zone to avoid.

\section{Results}

\subsection{Spectral Types, Radial Velocities, and Spectroscopic Binaries}

The spectral types for the sample, estimated from comparison
with the template stars shown in Figure 1, are given in column (7)
of Table 1 and range from K7 to M4.  Some
spectral types listed in Table 1 differ
from those found by Mart\'{i}n et al. (1998) (using moderate
resolution, visible light spectroscopy) by up to 2 spectral classes.
However, most typically agree to within 1 spectral class with
Mart\'{i}n et al.'s results.  Radial velocities were measured for
the targets by cross-correlating each spectrum of a particular target
against the closest matching template, rotationally broadened to the
best fit $v$sin$i$.  Rotational broadening of the template spectra
was accomplished using routines developed by C. Bender
(priv. comm.), following
Gray (1992) with the limb-darkening law of Claret (2000).
Both spectral type and estimated $v$sin$i$
were determined by visual inspection of the spectra and
assumed to apply to all the data for an individual object.  The 
cross-correlation peaks were fit with Gaussian functions to
determine the radial velocities.

The uncertainty in velocity measurements
depends in part on how well it is possible to measure the
radial velocities of the template standard stars with NIRSPEC.
For the template library
used here, the systematic uncertainty was estiamted at $\sim$1~km~s$^{-1}$
(Prato et al. 2002a; Mazeh et al. 2003).  Column (8) of Table~1
shows the average radial velocities measured
for each young star target in this sample and column (9) provides the rms
scatter for each object, $\sigma_{v_{rad}}$.  The average
$\sigma_{v_{rad}}$ for the 30 objects in Table~1 with more than
one velocity measurement, excluding the two obvious
SB2s, is 0.90~km~s$^{-1}$.  All objects with
$\sigma_{v_{rad}}$ greater than about five times this average are
defined as SBs.  I categorize one object with
$\sigma_{v_{rad}}=3.22$~km~s$^{-1}$
as radial velocity variable.  If the two SB1s, the radial velocity
variable, and ROXR1 31, for which the observed signal to noise
ratio was rather low for one epoch (see \S 4.3.6), are also removed from
the calculation, the average $\sigma_{v_{rad}}$ is 0.43~km~s$^{-1}$.
I therefore define the {\it internal} uncertainty in the radial
velocity variability measurements as 0.43~km~s$^{-1}$. 

In Table 1, each of the 33 targets observed individually for
spectroscopic multiplicity is listed separately.  There are two
SB2s, 2 SB1s, and one radial velocity variable.  The primary star
velocities for the SB1s and the radial velocity variable
appear in Table 3.  For the SBs, both the primary and secondary star
velocities, along with the mass ratios and center of mass
velocities, are given in Table 3 (see also \S 4.3).
The spectral types for the SB2s were derived from the spectra taken
during minimum radial velocity difference between the components.
This was the orbital configuration when the first spectrum  was
taken for both SB2s; thus, both initially appeared single.
Because the mass ratios are both close to unity, the spectral
types are are unlikely to differ by more than one spectral class.
Assuming that all of the newly discovered SB pairs presented here
have equal flux ratios, if angularly resolved,
none of the primary stars' magnitudes
would fall below $H=$11.1, the magnitude of the faintest object
in the survey.  Thus, there is no bias towards inclusion of
binaries in the sample based on brightness.

\subsection{Visual Binaries}

The excellent seeing often afforded on Mauna Kea revealed that a number
of the systems studied here have close visual companions, identified
by inspection of the SCAM images.  Greene \& Young's
(1992) study of infrared sources in the $\rho$~Oph core region showed
that the surface density of background {\it field stars}, appearing in the
same region as the young stars with K magnitudes of $<$13, is 0.07
per square arcminute, or 1.9$\times$10$^{-5}$ field stars per square
arcsecond.  For M stars, the $H-K$ colors are at most $\sim$0.3 mag, so
$K$-band data are an adequate surrogate for $H$-band observations.
Assuming a similar contamination surface density in all of the Ophiuchus
sub-regions I surveyed, there is a 72\% chance of one interloper
within 20$''$ of the 30 systems observed and an 18\% chance
of an interloper within 10$''$.  The probability of
a subarcsecond interloper is $<$1\% for the 30 systems
on the whole.

The contamination from the denser distribution of {\it young}
stars in the $\rho$~Oph core is 3.4 times greater than
from field stars.  Thus the chance for a 
superposition of a young star within 10$''$ of one of the target
stars is 63\%.  I therefore adopt a 10$''$
upper limit for a probable physical binary companion,
consistent with that used by Simon et al. (1995).  Given that
the R7, streamers, and smoke ring regions are probably less
dense than the $\rho$~Oph core region, this is likely an upper limit on
the young star contamination probability.

In total, I discovered five visual binary
systems with separations of $\sim$1$''$ or less, all
located in the smoke rings and R7.  For three of 
these pairs it was possible to take individual spectra of both
components (\S3), hence the 33 individual, unresolved objects surveyed
for radial velocity variability.  The separations of the other
two subarcsecond binaries were too small to obtain spectra of each
component without NIRSPEC behind the AO system (Comments, Table 1).  

There are two targets with companions between 1$''$ and 10$''$; in
one case a fainter, presumably later type star is $\sim$7$''$ away from
RX~J1613.8$-$1835, and in the other, an F star is located just
under 10$''$ from RX~1621.4$-$2332 (Vrba et al. 1993).  Four more systems 
have companions between 10$''$ and 20$''$.  Two of these are listed as
distinct targets.  The other two were
observed once but do not appear to be young M stars; one is
ScoPMS~52, a K0 star (Walter et al. 1994) and the other, $\sim$17$''$
west of RX~J1612.6$-$1924, has a featureless
spectrum.   One of the SB2s comprises one
component of a subarcsecond pair (RX~J1622.7-2325Nw); the other
component in this 0.9$''$ pair is an 0.1$''$ binary (RX~J1622.7-2325Ne),
revealed serendipitously with AO. The 1$''$ separation pair
RX~J1612.6-1924Ew$-$Ee harbors at least one radial velocity variable.  To
search for visual companions with separations smaller than $\sim$0.4$''$
it will be necessary to carefully observe this sample with AO;
initial AO snap-shot observations of about half the sample will
be discussed in a future paper.

\subsection{Comments on Individual Systems}

\subsubsection{RX~J1612.3$-$1909}

The four spectra of this SB1 system reveal rapid rotation
on the order of $v$sin$i\sim$40$-$50~km~s$^{-1}$.  As a result,
the determination of spectral type is challenging.  A broad,
boxy feature at $\sim$1.563~$\mu$m suggests that it is earlier
than the M2.5 designation of Walter et al. (1994) and 
Mart\'{i}n et al. (1998).  Figure 1 shows this spectral region
for the template standard stars.  An M2.5 spectrum, rotationally
broadened to 50~km~s$^{-1}$, would show a fairly triangular feature
at $\sim$1.563~$\mu$m.  I have therefore classified it as an M1.
A recent low-dispersion, visible light spectrum is consistent
with this classification (F. Walter, priv. comm.).

In addition to making the spectral type classification more difficult,
broad absorption lines can complicate cross-correlation
and yield less precise measurements of radial velocity.  The
four radial velocities measured for RX~J1628.0$-$2448 appear in
Table 3.  In order to investigate the possibility that their variability
is the result of poor measurement precision
resulting from broad spectral lines, I examined the cross-correlation
results for the other
$v$sin$i\sim$50~km~s$^{-1}$ target in the sample, RX~J1628.0$-$2448.
Cross-correlation of an M2 template spectrum,
rotated to 50~km~s$^{-1}$, against RX~J1628.0$-$2448 produced
the identical radial velocity for the
spectra from two epochs taken $\sim$1 year
apart.  This lends confidence to the  SB1 designation of
RX~J1612.3$-$1909; over the 14 months during which the 4 spectra
were taken the velocity varied by as much as 14~km~s$^{-1}$.

\subsubsection{RX~J1612.6$-$1924E}

About 17$''$ away to the west of RX~J1612.6$-$1924E is an
object with a featureless spectrum; however, the coordinates
listed under RX~J1612.6$-$1924 in SIMBAD point to that object.
Hence the eastern target is specified as ``E'' in Table 1 to
avoid confusion.  RX~J1612.6$-$1924E is itself a 1$''$ binary.
Both components of this visual pair were observed, although,
unfortunately, RX~J1612.6$-$1924Ee only once.
Assuming an uncertainty of 1~km~s$^{-1}$ for this single
radial velocity measurement, $\sim-$12~km~s$^{-1}$, indicates
that it is off from the sample average of $-$6.8~km~s$^{-1}$,
excluding only
the four SBs, by 5$\sigma$.  RX~J1612.6$-$1924Ew was observed 3 times
and shows a radial velocity rms in these measurements of 3.22~km~s$^{-1}$.
I therefore characterize Ew as a radial velocity variable.
The radial velocities of the two RX~J1612.6$-$1924E components should be
measured again to determine if one or both of these stars are SBs.

\subsubsection{RX~J1622.7$-$2325N}

An M0.5 star also lies
$\sim$13$''$ to the south of this system, designated as RX~J1622.7$-$2325S
in Table 1.  RX~J1622.7$-$2325N is an 0.9$''$ binary.  The
eastern component is an 0.1$''$ pair discovered serendipitously
in an AO observation in February, 2005, with the NIRC2 camera
on Keck~II.  RX~J1622.7$-$2325Nw is an SB2 with a mass ratio
of q$=$0.94$\pm$0.03 and a center of mass velocity
$\gamma=-$6.21$\pm$0.82~km~s$^{-1}$, consistent with the non-SB sample
average.  Figure 2 shows the spectra from the 4 epochs of observation.
The primary versus secondary radial velocities, easily determined from
the double peaks in the cross-correlation plots, are plotted in Figure 3
and tabulated in Table 3.  For the three epochs during which the two
stars in the SB showed distinct radial velocities, the correlation
peak of the primary was always slightly larger than that of the
secondary.
The mass ratio is the negative of the slope of the fit to these
points and the $\gamma$ velocity is equal to the y-intercept
divided by 1$+$q (Wilson 1941).  The uncertainties in the slope and
y-intercept were determined by a linear fit to the velocity data
taking into account the errors in both $v_1$ and $v_2$ (Table 3).
These uncertainties
were then propagated to determine the uncertainty in q and $\gamma$.

To determine an upper limit for the period of the system
I assumed component masses of 0.65~M$_{\odot}$ for each
star in the SB2 (Luhman 2000).  The maximum velocity amplitude 
measured for the primary star
was $\sim$51~km~s$^{-1}$.  Therefore, given q$=$0.94, the period must
be less than 11 days.  A less stringent estimate may also be made on
the basis of the spectra themselves.
The component radial velocities (Table 3) for the second and third epochs
of observation of RX~J1622.7$-$2325Nw were almost identical, implying
that the system had completed at least
one full orbit between those observations, taken 33 days apart.

\subsubsection{RX~J1622.8$-$2333}

Because the $v$sin$i$ of this SB1 is about 25~km~s$^{-1}$
it should be possible to analyze the spectra to identify the
secondary component and convert it into an SB2 (e.g., Prato
et al. 2002a).  However, with no a priori knowledge of the 
orbital elements, this will require more than the 4 observations
acquired as part of this work.  Application of
a two-dimensional cross-correlation algorithm, such as TODCOR
(Zucker \& Mazeh 1994) may also help to identify the secondary.
It is possible that either the mass ratio is
small or the inclination of the system is close to face on: no
evidence for lines of the secondary star were found in the 
one-dimensional cross-correlation analysis used.  The maximum
observed velocity variation was on the order of 9~km~s$^{-1}$
(Table 3).  The second and third epoch velocities are similar,
again providing an upper limit for the period of the system,
in this case 80 days.

\subsubsection{ROXR1 14}

ROXR1 14 is an SB2 with q$=$0.98$\pm$0.02 and
$\gamma=-$6.16$\pm0.43$~km~s$^{-1}$,
consistent with the non-SB sample average radial velocity.
Figure 4 shows the spectra from the 4 epochs of observation.
The primary versus secondary velocities, determined as for
RX~J1622.7$-$2325Nw, are shown in Figure 5, and tabulated
in Table 3.  The best
linear fit to the data, including the velocity errors, was used in the determination of q and $\gamma$ and their associated uncertainties.
As for RX~J1622.7$-$2325Nw, the correlation
peak of the primary was always slightly larger than that of the
secondary when separable.

Making the same assumptions regarding systemic mass as for
RX~J1622.7$-$2325Nw, i.e. M$_{tot}=1.3 M_{\odot}$,
combined with a maximum measured primary radial velocity amplitude
of 40~km~s$^{-1}$ and q$=$0.98, yields an upper limit for the period
of ROXR1 14 of 24 days.  There were no measurements of identical
velocities made close in time to provide additional information.

\subsubsection{ROXR1 31}

Despite the anomalous rms scatter in the 3 radial velocity
measurements of this system, it is unlikely that ROXR1 31 is
a radial velocity variable.  Two of the measurements, of
spectra with similarly high signal to noise ($\sim$200), were very close,
$-$7.01~km~s$^{-1}$ and $-$7.84~km~s$^{-1}$.  For the other
observation, made during deteriorating weather conditions, the
signal to noise is only $\sim$60 and the radial velocity 
measured is $-$4.13~km~s$^{-1}$.  Although a follow-up 
measurement of this target will be important to conclusively
rule out radial velocity variability, this is likely an example
of the loss of precision associated with poor signal to noise
observations.

\subsection{Cluster Velocities}

Excluding the four SBs, the radial velocity variable, and the
object observed only on one occasion (RX~J1612.6-1924Ee),
the average radial velocity of the sample is
$-$6.27~$\pm$1.48~km~s$^{-1}$.  This agrees well with the
average radial velocity, $-$6.3~$\pm$1.0~km~s$^{-1}$, found for
several $\rho$~Oph objects and about
a dozen nearby Upper Scorpius very low-mass young targets
(Kurosawa et al. 2006).  The average
radial velocities of the three sub-regions studied here, in which more
than 2 stars were observed, are
also statistically indistinguishable from the overall average.
This implies, as underscored in Kurosawa et al., that a single
velocity measurement that is more than 2$-$3~$\sigma$ from this mean
should be repeated to search for potential radial
velocity variables.  Indeed, RX~J1612.6-1924Ee may turn out to
be a spectroscopic binary and, on the basis of its discrepant
radial velocity, certainly merits follow-up observations (\S4.3.2).
However, both of the SB2s discovered here
were initially observed, coincidentally, at phases during which the
stellar motions were tangential to the line of sight.  At least
two radial velocity measurements may therefore be crucial.
Covey et al. (2006) measured the radial velocities of a sample of
Class I prototstars and compared these to the velocity of the
local CO gas.  They identify one candidate radial velocity
variable M star in Ophiuchus using this approach.

The constancy of the velocity dispersion in the areas around
the $\rho$~Oph core, as well as in the more distant smoke rings
region and in Upper Scorpius, provide input as to the nature of
star formation in this molecular cloud complex.  Feigelson (1996)
argues that this implies a slow and continuous rate of star
formation over a long period, after which thermal motions
disperse the members with a spread of $\sim$1~km~s$^{-1}$.
The Chamaeleon cloud presents a similar star formation history,
in contrast to that observed in the discreet bursts of star
formation in isolated cloudlets, as observed in the
Taurus-Auriga complex (Feigelson).

\section{Discussion}

\subsection{Spectroscopic Multiplicity and Comparison to Literature}

The spectroscopic multiplicity of the low-mass targets listed in Table 1
is 4/33, or $12.1\%^{+7.9\%}_{-3.6\%}$.  The uncertainties were
determined using a binomial probability distribution,
after Burgasser et al. (2003).  Including the radial velocity variable,
RX~J1612.6$-$1924Ew (\S 4.3), the
spectroscopic multiplicity is 5/33, or $15.2\%^{+8.1\%}_{-4.3\%}$.
Mathieu et al. (1989) found a similar SB fraction in their study
of weak-lined T~Tauri stars in the Taurus-Auriga, Scorpius-Ophiuchus,
and Corona Australis star-forming regions, 9\%$\pm$4\%, although
their sample
consists of primarily G and K spectral type systems.  Indeed, only
2 out of the 25 young SBs listed in Mathieu (1994) are M stars.
The short period SB frequency of field stars is
indistinguishable, 12\%$\pm$3\% (Abt \& Levy 1976; Morbey \& Griffin
1987; Abt 1987).  These samples described in the literature, however,
are not as narrowly defined as the sample in the present survey.
The range of spectral types, as determined herein, cover a relatively flat
distribution from K7 to M4 with a median value of M1, corresponding to
a mass range of about 0.3$-$0.7~M$_{\odot}$ for ages of a few Myr
(Baraffe et al. 1998; Luhman 2000).  Almost half of this sample
are M0 to M1 stars.

Early surveys of T~Tauri stars for SBs described by Mathieu et al. (1989)
and Mathieu (1994) considered higher mass objects, such
as the sample of Walter et al. (1988; Table 1), on average about
two times more massive than the sample described here.  Thus, although
recent results from the surveys of Joergens (2006) and Kurosawa
et al. (2006) have focussed mainly on samples of about a dozen
young substellar objects
(i.e. targets of spectral type later than M6) and a few very low
mass (spectral type M2.5$-$M5.5) stars, this paper describes
the largest homogeneous survey of low-mass T~Tauri stars to date.
The average spectral type of those presented in Table 1 is M1.
Three of the four SBs are systems with M1 star primaries; the
fourth has a K7 primary.  Although the occurrence of SBs is
greater therefore in the higher mass bins, it is consistent with
the distribution of masses studied.  It is unlikely that any SBs
were missed in the lower mass bins on the basis of poor radial
velocity precision as the scatter in radial velocity measurements
is indistinguishable between the early and late spectral type bins.

Melo (2003) draws attention to an excess in the frequency of spectroscopic
binaries in Ophiuchus compared to other small SFRs,
such as Taurus.  SBs were identified in the three sub-regions with
more than two stars surveyed, the smoke rings, R7, and L1688 core.
Surveys parallel to this one, including similar sample sizes
selected with the same criteria, will
help to verify a true differences between molecular cloud complexes.
Currently, the numbers of young stars in any given
region observed for spectroscopic multiplicity are small and
the samples irregular.  For example, observations by Melo (2003) of
about a dozen systems in Scorpius-Ophiuchus, 8 of them visual
binaries, did not angularly resolve the
components.  In some cases, double lines can
be detected if one or both components of a visual binary are SBs;
however, the blended light
from three or more stars can also disguise the signal from a low-mass
or long-period companion.

Reipurth et al. (2002) studied a sample of $\sim$100 young stars in
mainly southern SFRs and identified 2 SBs and 5 radial
velocity variables.  Those authors conclude that a contributing factor
to the low
spectroscopic multiplicity detected may be attributable to complications
arising from observations of the complex, classical T~Tauri star
spectra.  My sample contains 4 classical systems, none of which were
found to be spectroscopic binaries or even radial velocity variables.
However, the average rms radial velocity
scatter of these 4 targets, $\sim$0.6~km~s$^{-1}$, is fairly
consistent with that of the non-SB average:  $\sim$0.43~km~s$^{-1}$.
Therefore, it is possible that the low SB detection rate of
Reipurth et al. resulted from other factors as well, such as
the relatively blue wavelength range of their
observations, 3600$-$5200~\AA, subject to stronger
veiling from accretion processes in classical T~Tauri stars
than $H$-band observations.
The slightly lower resolution of their spectra, 20,000, might
also account for the lower detection rate.  It is also possible
that Ophiuchus is unusually rich in spectroscopic binaries
as noted above, and that few of the Reipurth et al. targets
besides Haro~1-14c were located in this region.

\subsection{Potential Limitations of Survey}

In order to estimate the sensitivity of this survey to small
mass ratio systems, I consider a test case in which a 5~km~s$^{-1}$
primary star amplitude was measured in a 3 year period, i.e. the
duration of this survey.  This amplitude is consistent with the largest
velocity change measured for radial velocity variable RX~J1612.6$-$1924Ew.
Assuming that one fourth of the orbit is traversed in 3 years, the total
orbital period is 12 years\footnote{A binary comprised of two young
M1 stars, with total mass 1.3~M$_{\odot}$ (Luhman 2000), and P$=$12
years would have an angular separation of $\sim$0.04$''$ at a distance
of 140~pc.  This separation is equivalent to the diffraction limit of the
Keck II telescope in the $H$-band and therefore provides a
convenient definition for a spectroscopic versus a visual binary.}. For an
SB with an edge-on, circular orbit, and given $v_1=5$~km~s$^{-1}$
and P$=$12 years, the present study is sensitive to early M star SBs with
mass ratios, q$=$M$_2$/M$_1$, down to 0.6 (Figure 6).  Observations
of shorter period systems, however, are sensitive to smaller values of q.
For a primary star mass of $\sim$0.6~M$_{\odot}$ and a mass
ratio of 0.1, the velocity modulation of the primary in an edge-on
circular orbit will only be detectable at 5~km~s$^{-1}$ for periods
of $\la$40 days, or 0.11 years.  This illustrates the strong bias
towards detection of unity mass ratio systems among longer period SBs.  
Mass ratio distributions of SBs over a wide range of periods will
reflect this bias; however, improved velocity precision will help to
alleviate this limitation.  The 5~km~s$^{-1}$ radial velocity amplitude
is also conservative; a 5$\sigma$ detection and the internal
radial velocity uncertainty of 0.43~km~s$^{-1}$ implies that
measuring a 2.5~km~s$^{-1}$ should be achievable 
using the simple techniques applied here.  With a 2.5~km~s$^{-1}$
primary velocity amplitude, a system
with a mass ratio of 0.1 and an $\sim$0.6~M$_{\odot}$ primary could be
detected with a period as long as about 310 days.

The star spots which are common and at times abundant on young,
active stars can affect radial velocity measurements (Saar \&
Donahue 1997).  One of
the most active weak-lined T~Tauri stars known, Parenago 1724,
manifests a star spot induced periodic radial velocity modulation
of $\sim$5~km~s$^{-1}$ (Neuhauser et al. 1998).  The SB1s and
SB2s identified in this sample have primary star velocity variations
of 9$-$80~km~s$^{-1}$.  Thus spot induced apparent radial velocity
modulation is probably not significant.



\section{Summary}

I have surveyed a homogeneous sample of young ($\sim$1~Myr) M stars in the
Ophiuchus SFR.  The goals of this program were
to identify the SB frequency
of the sample, measure mass ratios of SB2 pairs, and begin to search
for long-period systems suitable for high-angular resolution and
interferometric orbit mapping.  The following points summarize the
conclusions of this work:\\
(1) Including all of the spectroscopic and visual binary components
detected, I detected evidence for a total of 41 stars in this sample,
located in 30 systems.  33 individual targets were observed
spectroscopically to search for variable radial velocities.\\
(2) There are 2 SB2s, 2 SB1s, and at least 1 radial velocity variable.
An internal precision of 0.43~km~s$^{-1}$ was achieved.\\
(3) The average spectral type of the sample was M1; three of the
SB primaries are M1 stars and one is a K7.  The objects in the sample
are located in 4 sub-regions of Ophiuchus.  The average radial velocity of
the entire sample and of the individual sub-region objects is
indistinguishable, supporting a scenario for continuous star formation
across this molecular cloud complex.\\
(4) Counting the 4 identified SBs in 33 systems, the spectroscopic
binary frequency in this sample is $12\%^{+8.0\%}_{-3.5\%}$, similar
to that found by previous spectroscopic studies of higher mass
young stars as well as field stars.  Therefore, there appears to be
no dearth of low-mass, stellar SBs in Ophiuchus.\\
(5) The sample is rich in visual binary pairs; this work revealed
3 new visual binaries, all of separation $\le$1$''$.  A hierarchical
quadruple was also identified:  RX~J1622.7$-$2325N is an 0.9$''$
pair within which the eastern star is an 0.1$''$ binary and the
western one an SB2.\\
(6) Both the SB and visual binary frequencies measured here
are almost certainly incomplete.
Adaptive optics or $HST$ observations will reveal additional angularly
resolved components. Followup observations to identify long-period,
highly eccentric, and low-mass ratio SBs will also be important for
a more stringent completeness limit.  The coincidence that
the SB2s discovered in this modest survey were both initially observed
when their components were close to the gamma velocity of the system
underscores the importance of multiple observations and broad
phase coverage.  Inspection of the first epoch spectra alone did not
reveal obvious double-lined structure.
At least one additional epoch of identical spectroscopic
observations of this sample may reveal additional SBs and confirm
the radial velocity variable.  High angular resolution
imaging may detect more visual binaries.  Denser sampling is
required to determine the orbital elements of the new SBs discovered.
Properties of this rich, initially X-ray identified sample, such as
extinction, circumstellar dust, rotational velocities, and a more complete
discussion of visual binarity will be presented in a future paper.

\acknowledgements
These observations benefitted from the expert operating skills of
the Keck OAs, in particular C. Wilburn, G. Puniwai, C. Sorenson,
J. Aycock, S. Magee, and T. Stickel.  Excellent support, technical
and logistical, was provided by G. Hill, J. Lyke, P. Opoki, P. Amico,
and B. Schaefer.  I am grateful to M. Simon for a thorough reading
of a draft of this paper, to O. Franz for a final
vetting, and to T. Barman, D. Kirkpatrick, H. Roe, M. Simon, B. Skiff,
and F. Walter for advice and information.  I acknowledge helpful
encouragement early on in this project from E. Becklin, M. Rich.,
and I. McLean.  The latter in particular provided both the
financial support for the initial stages of this work as well as
the academic freedom that allowed me to develop it.
I thank the anonymous referee for thoughtful comments that helped
to improve the presentation of this paper.
NSF grant AST 04-44017 and awards from the Keck PI Data
Analysis Fund to L. Prato have financed the majority of this research.
I am grateful to the NASA Keck TAC for the adequate time allocations
to conduct and complete this multi-semester program.
This work made use of the SIMBAD reference database, the NASA
Astrophysics Data System, and the data products from the Two Micron All
Sky Survey, which is a joint project of the University of Massachusetts
and the Infrared Processing and Analysis Center/California Institute
of Technology, funded by the National Aeronautics and Space
Administration and the National Science Foundation.
Data presented herein were obtained at the W.M. Keck
Observatory from telescope time allocated to the National Aeronautics
and Space Administration through the agency's scientific partnership
with the California Institute of Technology and the University of
California. The Observatory was made possible by the generous
financial support of the W.M. Keck Foundation.
I recognize and acknowledge the
significant cultural role that the summit of Mauna Kea
plays within the indigenous Hawaiian community.  I am very 
grateful for the opportunity to conduct observations
from this special mountain.

\clearpage

\begin{deluxetable}{lllcrccccl} 
\tablecaption{Target List}
\rotate
\tabletypesize{\footnotesize}
\tablewidth{0pt}
\tablehead{
\colhead{  } & \colhead{  } & \colhead{  } & \colhead{PMS} &
\colhead{$H$\tablenotemark{a}} & \colhead{\# of}& \colhead{Sp} & 
\colhead{$<v_{rad}>$} & \colhead{$\sigma_{v_{rad}}$} & \colhead{  }\\ 
\colhead{Object} &
\colhead{$\alpha$ (J2000.0)} & \colhead{$\delta$ (J2000.0)} & 
\colhead{Class} & \colhead{(mag)} & \colhead{Obs} & \colhead{Type} & 
\colhead{(km s$^{-1}$)} & \colhead{(km s$^{-1}$)} & \colhead{Comments}
}
\startdata
\multicolumn{10}{c}{\underline{Smoke Rings}}\\
RX J1612.3$-$1909 & 16 12 20.9 & $-$19 09 05 & W & 9.9 & 4 & M1 & $-$10.52 & 6.30 & SB1; broad lines; aka ScoPMS 51\\
RX J1612.6$-$1859 & 16 12 39.2 & $-$18 59 28 & C & 9.5 & 3 & M1 & $-$5.64 & 0.63 & ScoPMS 52, a K0, is 19.2$''$ E \\
RX J1612.6$-$1924Ew\tablenotemark{b} & 16 12 41.2 & $-$19 24 18 & W & 9.1 & 3 & M0 & $-$9.49 & 3.22 & radial velocity variable \\
RX J1612.6$-$1924Ee\tablenotemark{c} & 16 12 41.2 & $-$19 24 18 & ... & ...& 1 & M1 & $-$11.97 & ... & Ew$+$Ee: 1$''$ pair\tablenotemark{d} \\
RX J1613.1$-$1904N & 16 13 10.2 & $-$19 04 13 & W & 10.0 & 3 & M2.5 & $-$5.50 & 0.24 & N: 0.5$''$ pair\tablenotemark{d}; spectra unresolved\\
RX J1613.1$-$1904S & 16 13 10.0 & $-$19 04 27 & C & 11.1 & 3 & M4 & $-$6.46 & 0.48 & \\
RX J1613.7$-$1926S & 16 13 43.9 & $-$19 26 49 & W & 9.2 & 3 & M0 & $-$6.88 & 0.24 & \\
RX J1613.7$-$1926N & 16 13 43.9 & $-$19 26 49 & ... & ... & 3 & M1 & $-$8.39 & 0.24 & S$+$N: 0.7$''$ pair\tablenotemark{d} \\
RX J1613.8$-$1835 & 16 13 47.5 & $-$18 35 01 & W & 10.3 & 3 & M1.5 & $-$6.19 & 0.41 & fainter, unobserved star 7.3$''$ N \\
RX J1613.9$-$1848 & 16 13 58.2 & $-$18 48 29 & W & 10.1 & 4 & M2 & $-$6.91 & 0.52 & \\
RX J1614.4$-$1857 & 16 14 28.9 & $-$18 57 23 & W & 9.7 & 3 & M2.5 & $-$5.78 & 0.41 & \\
RX J1615.1$-$1851 & 16 15 08.6 & $-$18 51 01 & W & 9.9 & 3 & M0 & $-$4.95 & 0.41 & \\
\multicolumn{10}{c}{\underline{R7}}\\
RX J1621.4$-$2332 & 16 21 28.7 & $-$23 32 49 & W & 10.1 & 3 & M0.5 & $-$7.15 & 0.48 & brighter, unobserved star 9.7$''$ N\tablenotemark{e} \\
RX J1622.6$-$2345 & 16 22 37.6 & $-$23 45 51 & W & 10.1 & 3 & M2 & $-$6.46 & 0.48 & \\
RX J1622.7$-$2325Ne & 16 22 46.8 & $-$23 25 33 & W & 8.7 & 4 & M0.5& $-$5.78 & 0.75 & Ne: 0.1$''$ pair\tablenotemark{d}\\
RX J1622.7$-$2325Nw & 16 22 46.8 & $-$23 25 33 & ... & ... & 4 & M1 & ... & ... & SB2; Ne$+$Nw: 0.9$''$ pair\tablenotemark{d}\\ 
RX J1622.7$-$2325S & 16 22 47.2 & $-$23 25 45 & W & 10.0 & 3 & M0.5 & $-$6.88 & 0.63 & \\
RX J1622.8$-$2333 & 16 22 53.4 & $-$23 33 10 & W & 9.9 & 4 & K7 & $-$3.40 & 4.41  & SB1\\
RX J1622.9$-$2326 & 16 22 59.9 & $-$23 26 35 & P & 9.4 & 3 & M0 & $-$7.29 & 0.24 & \\
\multicolumn{10}{c}{\underline{$\rho$ Oph Core}}\\
RX J1624.2$-$2427 & 16 24 15.9 & $-$24 27 35 & P & 9.6 & 3 & M0 & $-$5.91 & 0.48 & \\
RX J1625.2$-$2455 & 16 25 14.7 & $-$24 55 45 & W & 8.6 & 3 & K7 & $-$6.88 & 0.95 & \\
ROXR1 3 & 16 25 24.3 & $-$24 29 44 & C & 9.2 & 4 & M3 & $-$6.60 & 0.58 & \\ 
ROXR1 7 & 16 25 47.7 & $-$24 37 39 & W & 10.8 & 3 & M2.5 & $-$3.99 & 0.48 & \\
ROXR1 14 & 16 26 03.3 & $-$24 17 47 & W & 9.6 & 4 & M1 & ... & ... & SB2\\
RX J1626.3$-$2407 & 16 26 18.8 & $-$24 07 19 & W & 9.8 & 3 & M2.5 & $-$7.29 & 0.63 & \\
ROXR1 20 & 16 26 19.5 & $-$24 37 27 & W & 10.4 & 3 & M4 & $-$6.46 & 0.24 & \\
ROXR1 31 & 16 26 44.3 & $-$24 43 14 & W & 9.9 & 3 & M1.5 & $-$6.33 & 1.95\tablenotemark{f} & \\
RX J1627.6$-$2404W & 16 27 38.3 & $-$24 04 01 & W & 9.5 & 3 & K7 & $-$7.01\tablenotemark{g} & ... & \\
ROXR1 51 & 16 27 39.4 & $-$24 39 16 & C & 9.3 & 3 & M0 & $-$6.60 & 0.83 & \\
RX J1627.6$-$2404E & 16 27 41.9 & $-$24 04 27 & P & 10.0 & 3 & M2 & $-$7.29 & 0.48 & \\
RX J1628.0$-$2448 & 16 28 00.0 & $-$24 48 19 & W & 9.8 & 2 & M2 & $-$9.49\tablenotemark{g} & ... & broad lines\\
\multicolumn{10}{c}{\underline{Streamers}}\\
RX J1632.7$-$2332 & 16 32 44.4 & $-$23 32 13 & P & 10.9 & 3 & M2.5 & $-$9.63 & 0.24 & \\
RX J1636.2$-$2420 & 16 36 16.9 & $-$24 20 34 & P & 10.3 & 3 & M3 & $-$2.06\tablenotemark{g} & ... & \\
\enddata

\tablecomments{The best coordinates were defined using $2MASS$ positions of the targets.  In most
cases these were consistent with SIMBAD, however, for about 20\% of the targets, the
SIMBAD coordinates were off by 6$''$ or more.}

\tablenotetext{a}{$H$ magnitudes from $2MASS$; for the three sub-arsecond visual binaries with both
components listed the total systemic magnitude is given with the entry for the primary.}
\tablenotetext{b}{SIMBAD coordinates for the Ew-Ee pair point to an object with a featureless spectrum
17.2$''$ to the west.}
\tablenotetext{c}{The eastern component in this 1$''$ pair was only observed once but has a velocity
that is distinct from the local mean; it should be reobserved for variability.}
\tablenotetext{d}{This work.}
\tablenotetext{e}{Vrba et al. (1993) classify this object, VSS II-137, as an F star.}
\tablenotetext{f}{Radial velocity scatter probably attributable to poor SNR, 60, in one spectrum vs. 200 in the others, but should be reobserved.}
\tablenotetext{g}{All radial velocity measurements equal.}

\end{deluxetable}

\clearpage

\begin{deluxetable}{llll} 
\tablecaption{Spectral Type Standards}
\tablewidth{0pt}
\tablehead{
\colhead{Object} & \colhead{Spectral Type} & \colhead{  } &
\colhead{Spectral Type}\\
\colhead{Name} & \colhead{from literature} & \colhead{Reference} &
\colhead{determined here}
}
\startdata
GJ 763 & M0 & 1  & M0 \\
GJ 752A & M3 & 1 & M2 \\
GJ 436 & M3 & 1 & M2.5 \\
GJ 15A & M1.5 & 2 & M3 \\
GJ 402 & M4 & 1 & M4 \\
\enddata


\tablerefs{
(1) Kirkpatrick et al. (1991);
(2) Henry et al. (1994).}

\end{deluxetable}

\begin{deluxetable}{ccc}
\tablewidth{0pt}
\tablecaption{Spectroscopic Binary Component Velocities\label{tbl-3}}
\tablehead{
\colhead{BJD\tablenotemark{a}} & \colhead{$v_1$\tablenotemark{b}} & \colhead{$v_2$\tablenotemark{b}} \\
\colhead{(2,450,000$+$)} & \colhead{(km s$^{-1}$)} & \colhead{(km s$^{-1}$)}}
\startdata
\multicolumn{3}{c}{\underline{RX J1612.3$-$1909}}\\
  2362.99      & $-$14.44 & ... \\
  2447.76	& $-$2.06 & ... \\
  2723.01	& $-$9.49 & ... \\
  2773.86	& $-$16.09 & ... \\
\tableline
\multicolumn{3}{c}{\underline{RX J1612.6$-$1924Ew\tablenotemark{c}}}\\
  2723.06	& $-$13.20 & ... \\
  2773.99	& $-$7.43  & ... \\
  3150.86	& $-$7.84  & ... \\
\tableline
\multicolumn{3}{c}{\underline{RX J1622.7$-$2325Nw,~~q$=$0.94$\pm$0.03,~~$\gamma=-6.21\pm$0.82~km~s$^{-1}$}}\\  
  2449.80      & $-$6.19 & $-$6.19 \\
  2771.99	& 44.98 & $-$61.90 \\
  2804.94	& 44.56 & $-$60.66 \\
  3198.79	& 31.77 & $-$47.87 \\
\tableline
\multicolumn{3}{c}{\underline{RX J1622.8$-$2333}}   \\
  2473.82      & $-$9.49 & ... \\ 
  2724.06      & $-$0.41 & ...  \\ 
  2804.97      & $-$0.83 & ... \\ 
  3199.79	& $-$3.71 & ... \\
\tableline
\multicolumn{3}{c}{\underline{ROXR1 14,~~q$=$0.98$\pm$0.02,~~$\gamma=-6.16\pm$0.43~km~s$^{-1}$}}   \\
  2363.16      & $-$6.19  &  $-$6.19 \\ 
  2449.90      & 34.25  & $-$47.45   \\ 
  2773.89      & $-$42.50    & 30.95 \\ 
  3124.97      & 33.01    & $-$46.22 \\
\enddata

\tablenotetext{a}{Barycentric Julian date.}
\tablenotetext{b}{Systematic uncertainties are $\sim$1~km~s$^{-1}$.}
\tablenotetext{c}{Radial velocity variable; see Table 1 and \S4.3.2.}

\end{deluxetable}

\clearpage

\begin{figure*}
\includegraphics[angle=0,width=6.0in]{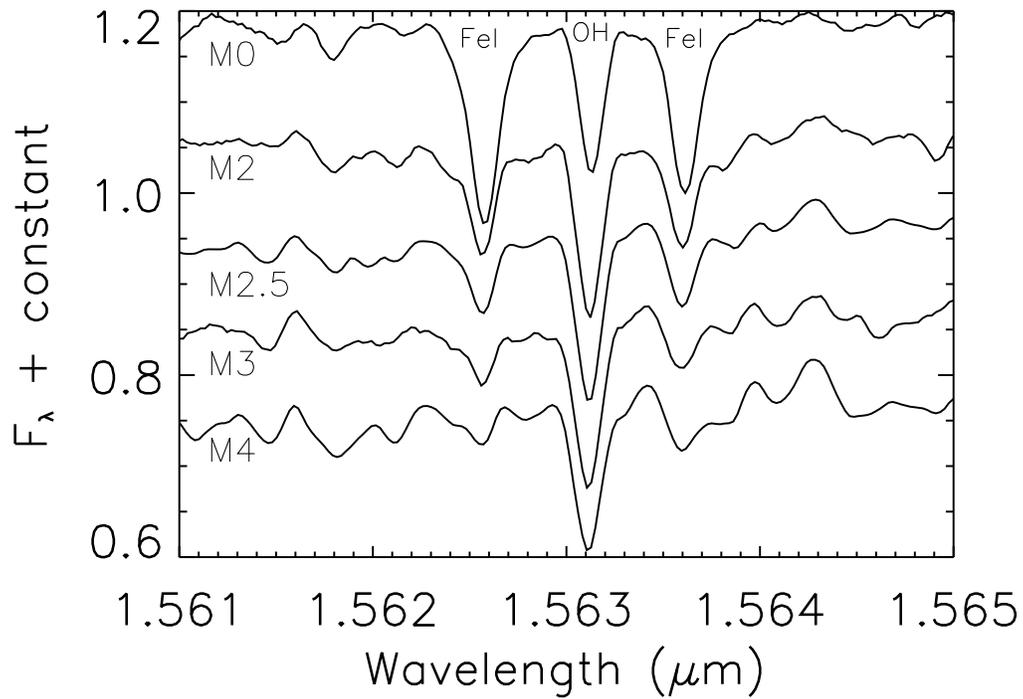}
\caption{Spectral type standard star sequence, as determined from
high-resolution $H$-band spectra.  See Table 2 for spectral types
identified in the literature.  All spectra are presented on a
heliocentric wavelength scale and all have been rotationally
broadened to $v$sin$i\sim$15~km s$^{-1}$, close to the average
$v$sin$i$ for the sample.}
\label{fig:mstrs}
\end{figure*}

\begin{figure*}
\includegraphics[angle=0,width=5.5in]{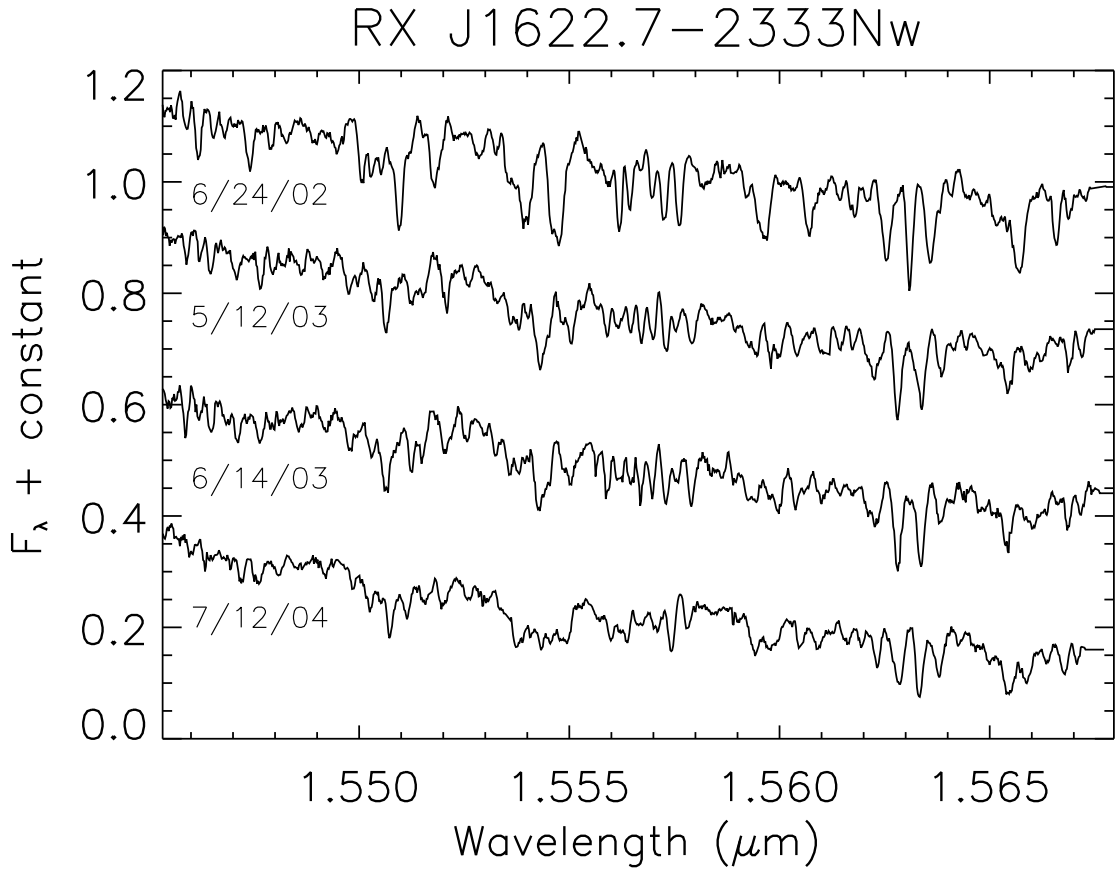}
\caption{Four epochs of spectra for the SB2 RX~J1622.7$-$2325Nw;
UT dates of the observations are indicated.}
\label{fig:SB2_1622.7}
\end{figure*}

\begin{figure*}
\includegraphics[angle=0,width=5.5in]{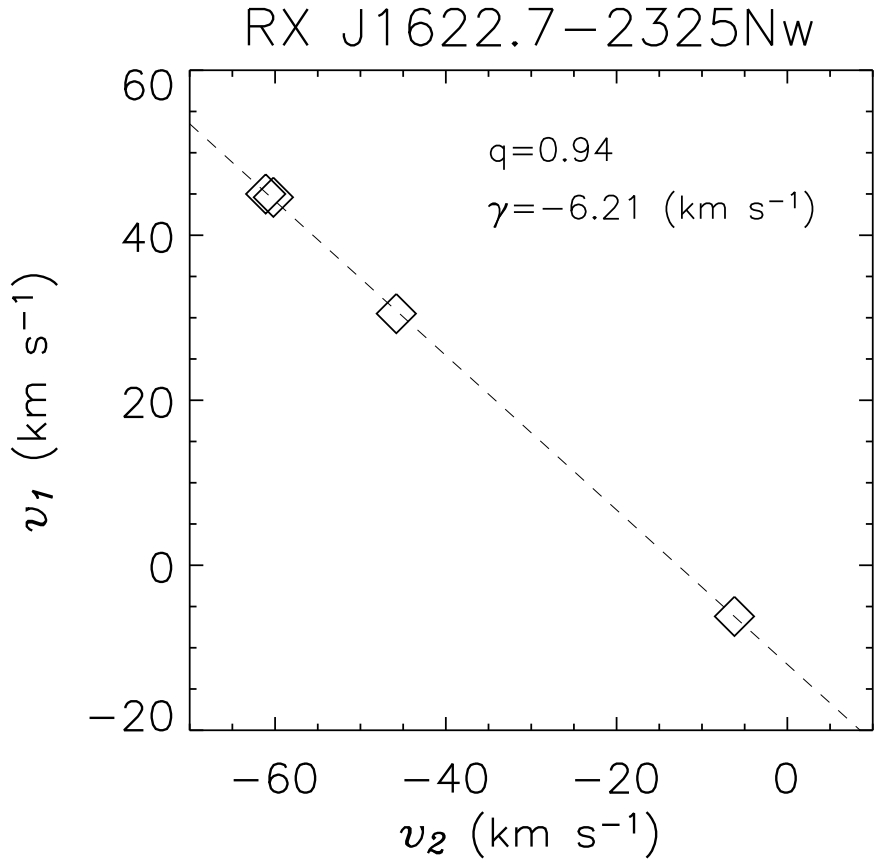}
\caption{Linear fit to the primary versus secondary
radial velocities for RX~J1622.7$-$2325Nw.}
\label{fig:W41_rxj}
\end{figure*}

\begin{figure*}
\includegraphics[angle=0,width=5.5in]{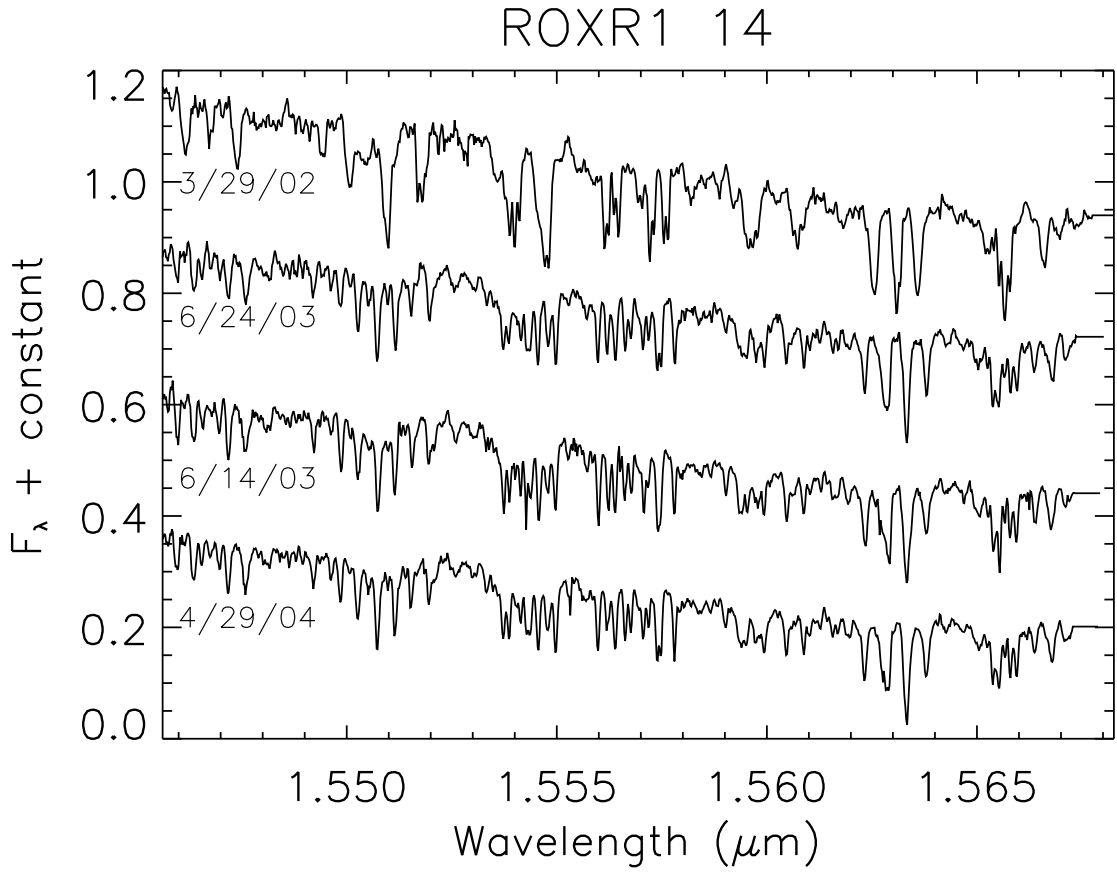}
\caption{Four epochs of spectra for the SB2 ROXR1 14;
UT dates of the observations are indicated.}
\label{fig:SB2_ROXR1_14}
\end{figure*}

\begin{figure*}
\includegraphics[angle=0,width=5.5in]{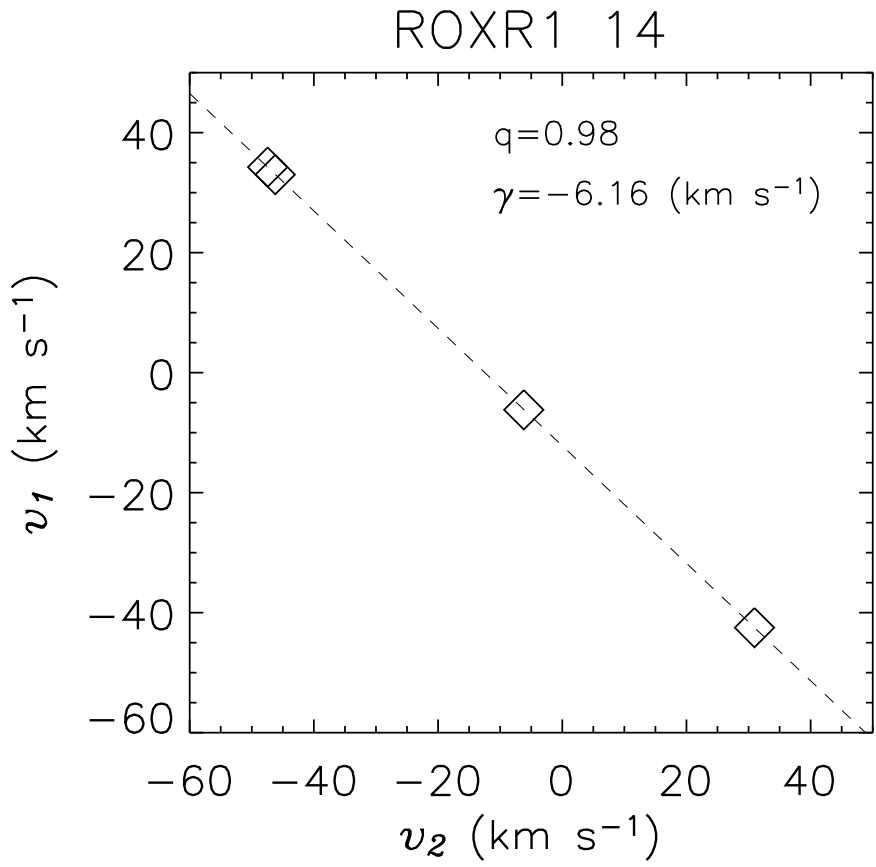}
\caption{Linear fit to the primary versus secondary
radial velocities for ROXR1 14.}
\label{fig:W41_rox}
\end{figure*}

\begin{figure*}
\includegraphics[angle=0,width=6.0in]{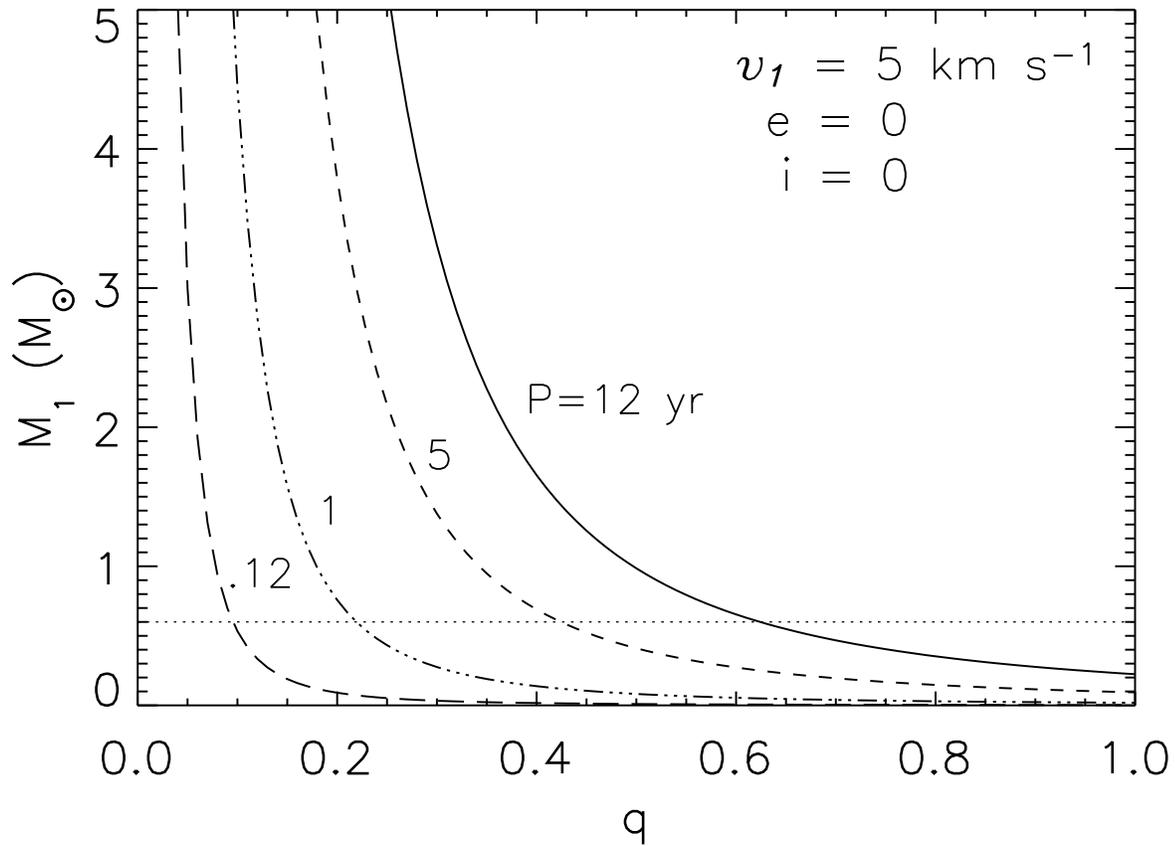}
\caption{Primary star mass as a function of mass ratio, q, for
a range of orbital periods from 0.12 to 12 years and a primary
star velocity of 5~km s$^{-1}$.  Assumes circular, edge-on orbits.
To detect a mass ratio of 0.1 in a system with a primary star
mass of 0.6~M$_{\odot}$, implying a substellar mass companion
(dotted line), requires a period of less than 0.11 years,
or $\sim$40 days.}
\label{fig:prds}
\end{figure*}

\end{document}